\documentclass{epl}
\title{\bf Off-equilibrium dynamics of the two-dimensional
Coulomb glass}
\author{D. R. Grempel}
\institute{CEA-Saclay, DSM/DRECAM/SPCSI, 91191 Gif-sur-Yvette,
France}
\begin{document}
\maketitle
\begin{abstract}
The dynamics of the 2D Coulomb glass model is
investigated by kinetic Monte Carlo simulation. An exponential
divergence of the relaxation time signals a zero-temperature
freezing transition. At low temperatures the dynamics of the
system is glassy. The local charge correlations and the response
to perturbations of the local potential show aging. The dynamics
of formation of the Coulomb gap is slow and the density of states
at the Fermi level decays in time as a power law. The relevance of
these findings for recent transport experiments in
Anderson-insulating films is pointed out.
\end{abstract}

A low-temperature glassy phase that had been anticipated by
several authors~\cite{davies,pollak1} was recently observed in
studies of hopping conduction in disordered InO
films~\cite{films:experiments,films:aging}. The experiments show
sluggish, non-exponential, relaxation of the
conductance~\cite{films:experiments} as well as aging and memory
effects~\cite{films:aging} similar to those observed in
glasses~\cite{review:glasses1}. The details of the dependence of
the observed effects on carrier density and sample conductance
suggests that the glassiness of these systems is associated to the
interplay of disorder and electron
interactions~\cite{films:aging}.

The experimental systems, with very low carrier densities and very
large resistances, are deep in the Anderson-Mott localized phase.
In this regime quantum effects are qualitatively unimportant and a
description based on the classical 2D Coulomb glass model is
appropriate. The Hamiltonian of the model
 is~\cite{coulomb-glass}
\begin{equation}\label{hamil}
H=\sum_i n_i \varphi_i + {e^2 \over 2 \kappa} \sum_{i\ne j} {(n_i
- K) (n_j - K) \over \left|\vec{r}_i - \vec{r}_j\right|}\;.
\end{equation}
Here, $\vec{r}_i$ denotes the position of the $i$-th localized
state in the plane of the sample, $\varphi_i$ its energy, $e$ the
electron charge and $\kappa$ the background dielectric constant.
The electron occupancy $n_i=0,1$, and there is a uniform compensating
background charge density  $K=1/N\sum_i n_i $ that ensures global
charge neutrality. The
presence of the long range Coulomb interaction  in (\ref{hamil})
is a salient feature of this model that has strong effects on its
low-temperature properties.

The Coulomb glass model has been intensively investigated
by Monte Carlo simulation but so far the main emphasis has been on
its low-temperature properties at
equilibrium~\cite{davies,lee,yu1,rusos,efros}.
In this Letter we report the results of the first systematic investigation
of its off-equilibrium dynamics.

The time evolution of Anderson insulators takes place through
phonon-assisted processes in which electrons hop from occupied
sites
 $a$ to empty sites $b$ with a transition rate~\cite{coulomb-glass}
\begin{eqnarray}\label{rate}
\Gamma_{a\rightarrow b}& =& \tau_0^{-1} \exp\left({-{2 R_{ab}
\over \ell}}\right)\;\Phi\left({\Delta E_{ab}\over T}\right)
\;,\\
\label{energies} \Delta E_{ab}& =& \epsilon_b - \epsilon_a - {e^2
\over \kappa R_{ab}}\;,\;\;\;\epsilon_a =\varphi_a +  {e^2 \over
\kappa} \sum_{b\ne a} {\delta n_b  \over R_{ab}}\;.
\end{eqnarray}
Here, $\tau_0$ is a microscopic timescale, $\ell$ is the spatial
extension of the localized wavefunctions, $\delta n_a \equiv n_a -
K$ and $\Delta E_{ab}$ is the difference in energy between the final
and initial states. The origin of the exponential factor in Eq.~(\ref{rate})
is the  dependence of the electron-phonon tunneling matrix
 element between
sites $a$ and $b$ on the distance between them and $\Phi(x)$ is
the thermal factor, $\Phi(x)=e^{-x}$ for $x > 0$, $\Phi(x)=1$ for
$x < 0$. The positions $r_i$ of the localized states and their
energies $\varphi_i$ are both random in general. However, we
follow previous authors~\cite{lee,yu1} and set $\varphi_i=0$ thus
limiting our discussion to
 a simplified model with positional disorder only
 ~\footnote{The symmetry
breaking field $\varphi_i$ changes 
the universality class of the problem but the qualitative
 features of the
off-equilibrium dynamics are preserved,  
at least if the  random local potential is small enough. 
See also our concluding
remarks}. 
Furthermore, we take
$K=1/2$ endowing the model with particle-hole symmetry, a feature
that greatly simplifies the analysis of the
results~\cite{lee,yu1}. The substitution $2 n_i -1 \equiv \sigma_i
= \pm 1$ makes Hamiltonian (\ref{hamil}) formally equivalent to
that of a classical disordered antiferromagnet with long range
$1/R$ interactions. The dynamics (\ref{rate}) would be quite
unusual in the magnetic context,
however~\footnote{Eq.~(\ref{rate}) is a  widely used
 approximate form for
the transition rate in hopping-conduction systems. We expect other
forms for $\Gamma_{a\rightarrow b}$ to lead to a qualitatively similar
dynamics.}.

We measure distances and energies in units of the average
intersite distance $a_0$ and the Coulomb energy $E_C = e^2/(\kappa
a_0)$ respectively, and
we fix $\ell =a_0$. Using parameters appropriate for InO this
 corresponds to a
carrier density $n \sim 3\times 10^{21}$ {\rm cm}$^{-3}$ which is
within the range of values studied
experimentally~\cite{films:experiments,films:aging}. For these
values of the parameters
 $E_C \approx $ 0.1 eV is huge compared to the
experimental temperatures.

We performed kinetic Monte Carlo simulations of model
(\ref{hamil}). The localization centers
 are randomly distributed within a simulation cell of side $L$
containing $N_e=L^2/2$ electrons.  The cell is then periodically
repeated in 2D space. The contribution  to the potential within
 the simulation cell of charges located
outside it is computed using the Ewald method.
 We start the simulations from a random electron configuration
to mimic a quench from
 high temperature and then let the system evolve at the working
temperature $T$ with the dynamics~(\ref{rate}). The latter is
implemented as follows.  An occupied site $a$ is picked at random
and a probability $P\left(b \left|\right. a\right)\propto
\exp\left(-2 R_{ab}/\ell\right)$
 is assigned to each of the available empty sites $b$.
A destination site is then picked from this probability
distribution and an attempt is made to move the electron on site
$a$ to $b$. The
  move is accepted with probability $\Phi(\Delta E_{ab}/T)$.
A Monte Carlo step (MCS)
 consists of $N$ such attempts.
We simulated systems with $L=$8, 16, 32 and 64 in  the
temperature range $0.01 \le T \le 0.1$. The length of our runs was
typically $2\times 10^6$ MCS. Physical quantities were monitored
as a function of time and the results were averaged over
 between 64 and 6400 realizations of the disorder and the initial conditions,
depending on size and temperature.

We begin our discussion by considering
 the local charge correlation function,
\begin{equation}
C(t + t_w,t_w) = {4 \over N} \sum_i \langle \delta n_i(t+ t_w)\;
\delta n_i(t_w)\rangle\;, \label{C}
\end{equation}
where the brackets denote the average over configurational
disorder and thermal noise and the waiting time $t_w$
 is defined as the time elapsed since the quench. If $t_w$ is longer than the
equilibration time $\tau_{eq}$ there is time-translation
invariance and the correlation function will depend on the time
difference $t$ only. Otherwise, $C$ is expected to depend on both $t$
and $t_w$.
\begin{figure}
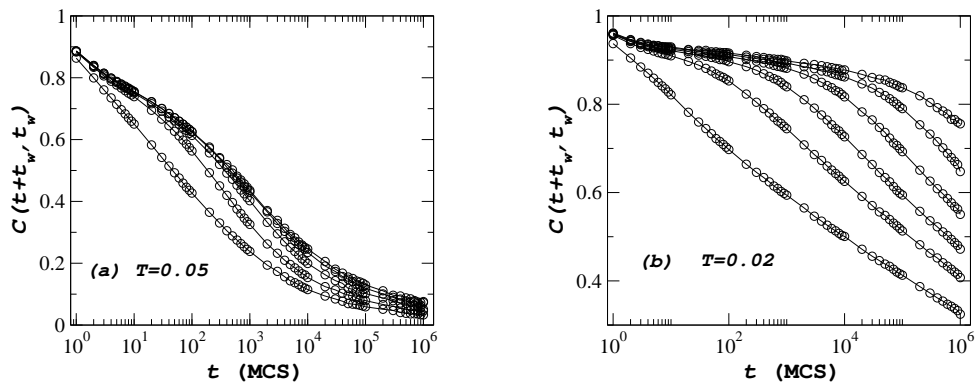

\twoimages[height=5cm]{fig1a.eps}{fig1b.eps} 
\caption{$C(t+t_w,t_w)$ for~$t_w = 10,10^2, 10^3, 10^4, 10^5$ and
$10^6$, from bottom to top, for $T=0.05$ (a) and $T=0.02$ (b).
} \label{fig1}
\end{figure}

The $T$-dependence of $\tau_{eq}$ was determined using the
method of Bhatt and Young~\cite{bhatt}. This is based on the
 comparison of two
overlap functions, $Q_{rep}(t_0)=\langle \left[4/N\sum_i
\delta n_i^a(t_0)
 \delta n_i^b(t_0)\right]^2\rangle$ and
$Q_{t}(t_0)=\langle \left[4/N\sum_i \delta n_i^a(2 t_0)
 \delta n_i^a(t_0) \right]^2\rangle$. In these expressions
the superscripts $a$ and $b$
 denote two copies of the system  ({\it i.e.}, two systems
with the same realization of the disorder) that evolve
independently starting
 from different uncorrelated initial conditions.
 If $t_0 \ge \tau_{eq}$ the system is at equilibrium in which
 case~\cite{bhatt}
 $Q_{ rep}(t_0)=Q_{t}(t_0)$. The equilibration
time may thus be defined as the value of $t_0$ at which the two curves
 merge (within the statistical uncertainty).
Fig.~\ref{fig6} shows the time-dependence of these overlap
functions for a system of size $N=256$ and several temperatures.
From a quantitative analysis of these results we find [cf. inset
to the figure] that
 $\tau_{eq}(T) \sim \exp(T_0/T)$
where, restoring units, $T_0 \sim E_C/2$. The exponential divergence
of $\tau_{eq}$ at $T=0$ is consistent with a freezing
transition at $T_g=0$ as reported by other
authors~\cite{lee,rusos}.
\begin{figure}
\twofigures[height=4.5cm]{fig6.eps}{fig5.eps} \caption{The overlap
functions $Q_{rep}$ (open symbols) and $Q_{t}$
(filled symbols). The curves have been displaced vertically for
the sake of clarity. The temperatures are 0.05, 0.06, 0.07, 0.08
and 0.09, form bottom to top. Inset : the $T$-dependence of the
equilibration time in MCS.}
\label{fig6} \caption{Size-dependence of $C(t+t_w,t_w)$ for
$T=0.03$, $t_w=10^5$ and $L$=8,16 and 32. Inset: time-dependence
of the domain size $\xi$ determined from the analysis of the four
point correlation function at $T=0.03$.} \label{fig5}
\end{figure}
The equilibration time is thus finite for all $T \ne 0$ but it
becomes very long for $T \ll E_C$. For $T > 0.05$ our observation
time is longer than $\tau_{eq}$ but the latter far exceeds the
simulation time for $T < 0.04$; in this  $T$-range we can only
probe the off-equilibrium dynamics. These temperature regimes are
illustrated in Fig.~\ref{fig1} that shows the auto-correlation
function for two temperatures, $T=0.05$ and $T=0.02$, and several
values of $t_w$. At the higher temperature the relaxation curves
for different values of $t_w$ coincide for
 $t_w > 10^5 $ MCS $ \sim \tau_{eq}$
indicating that equilibrium was attained
 in the course of the simulation.
At the lower temperature, however, the $t_w$-dependence persists throughout
the observation time and the relaxation becomes increasingly
sluggish as the system gets older. Equilibrium was obviously not
reached within the simulation time in this case.

The size-dependence of our results in the non-stationary regime is
remarkably weak as shown in Fig.~\ref{fig5}. This is the signature
of a very slow growth of the size of ordered domains $\xi(t)$
after a quench, a familiar feature of off-equilibrium glassy
systems. An analysis of the four-point correlation function
 ${\cal C}^{(4)}(R_{ij}) = \langle \delta n_i^a
\delta n_i^b n_j^a \delta n_j^b  \rangle $  along the
lines of Reference~\cite{rieger}
 gives access to $\xi(t)$
\footnote{Here we use the simple estimate $\xi^2(t) = \sum_{i,j}
R^2_{ij} {\cal C}^{(4)}(R_{ij})/\sum_{i,j}
 {\cal C}^{(4)}(R_{ij}).$}. As illustrated in the
inset to Fig.~\ref{fig5}, at low temperature, $\xi(t)
\propto \log t$ reaches only a few lattice spacings in the course
of our simulations.

Let us discuss in more detail the off-equilibrium regime
illustrated by the data of Fig.~\ref{fig1}~(b). At equilibrium, the
correlation function
 has the scaling form $C(t)\;\sim\;t^{-\lambda}\;{\cal
C}_{\rm eq}(t/\tau)$ with $T$-dependent $\lambda$ and $\tau$. In
the non-stationary regime this expression generalizes
to~\cite{rieger,godreche}
\begin{eqnarray}
C(t+t_w,t_w) \sim t^{-\lambda}\; {\cal C}_{\rm ag}\left({h(t+t_w)
\over h(t_w)}\right)\;. \label{mult-aging}
\end{eqnarray}
The above form for the argument of the scaling function ${\cal
C}_{\rm ag}$, often used in discussions of
 experimental data in
glasses~\cite{review:glasses1}, also arises in the exact solution
of certain mean-field models~\cite{CuKu}. Here we adopt the widely
used parameterization~\cite{review:glasses1}
 $h(x)=\exp\left[x^{1-\mu}/(1-\mu)\right]$.
In the limit $\mu \to 1$ this expression leads to
 simple aging, $C(t+t_w,t_w) \sim t^{-\lambda}\;
{\cal C}_{\rm ag}(t/t_w)$; for other values of $\mu$ it leads to
an effective relaxation increasing with age as $t_w^\mu$. We used
Eq.~(\ref{mult-aging}) to perform a scaling analysis of the data
of Fig.~\ref{fig1}~(b) as well as those obtained for
 two other temperatures,
$T=0.01$ and $T=0.03$. The results are displayed in
Fig.~\ref{fig2}. We see that Eq.~(\ref{mult-aging}) leads
to an excellent collapse of the scaled data for the three
temperatures. The exponent $\lambda$, small in magnitude, varies
substantially with temperature. The values obtained are consistent
with $\lim_{T\to 0}\lambda(T)=0$ as expected for system that freezes
at $T=0$. The variation of the parameter $\mu$ with $T$ is much
less pronounced. We find simple aging ($\mu \to 1$) as $T\to 0$
and a slight tendency to sub-aging with increasing $T$.
It is worth noticing that simple
aging was observed in the conductivity of Anderson-insulating
films~\cite{films:aging}.
\begin{figure}
\twofigures[height=5cm]{fig2.eps}{fig7.eps} \caption{Scaling plot of
$C(t+t_w,t_w)$. The symbols represent the data obtained for three
temperatures, $T=0.01, 0.02$ and $0.03$. Inset: the $T$-dependence
of the exponents $\mu$ and $\lambda$. }
\label{fig2} \caption{Response to a random local field of amplitude
$\varphi_0=0.01$ for $T=0.02$.}
 \label{fig7}
\end{figure}

It has been recently suggested~\cite{zvi-new} that some of the
experimentally observed features may be due to a slow response of
the system to the changes in the local random potential
$\varphi_i$ [cf. Eq.~(\ref{hamil})]. We thus studied the response
of the system to random perturbations of the form $\delta
\varphi_i = \epsilon_i \varphi_0$ where $\varphi_0 \ll 1$ is the
overall scale of the perturbation and $\epsilon_i$ are normalized 
random variables uncorrelated from site to site, $\langle \epsilon_i
\epsilon_j \rangle = \delta_{ij}$. 
This perturbation is
switched on at time $t_w$ and its effect on the system is observed
a time $t$ later. The quantity conjugated to the random potential
is $\delta n(t) \equiv 1/ N\; \sum_i\;\langle\delta n_i(t)
\epsilon_i\rangle$. The results of our measurements of the
response of the system are shown in Fig.~(\ref{fig7}) for the case
of a random potential of amplitude $\varphi_0=0.01$ and $T=0.02$.
It may be seen that the response $\delta n$ exhibits aging just as
the correlation function $C$ does. As $t_w$ increases, the
response gets more and more sluggish and, within the time window
explored, $\delta n(t) \approx \ln t$ for $t \gg t_w$.

\begin{figure}
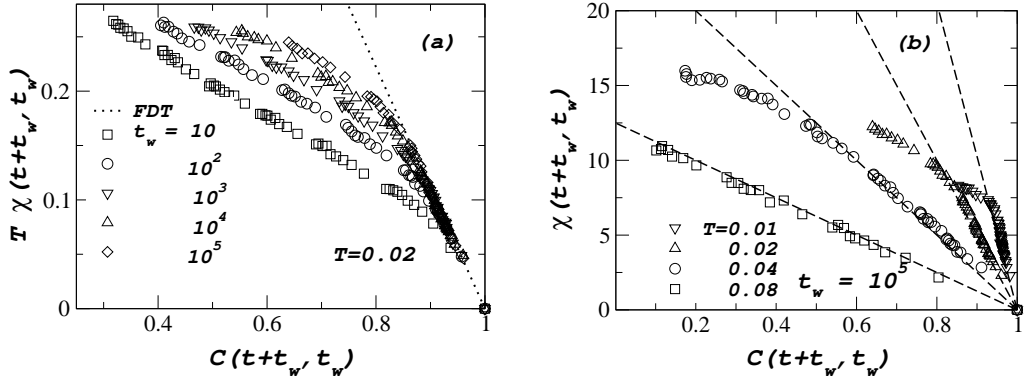

\twoimages[height=5cm]{fig3a.eps}{fig3b.eps}
 \caption{
(a) Parametric plot of susceptibility {\it vs.} correlation for
$T=0.02$ and the waiting times indicated in the legend.  (b)
$\chi$ {\it vs.} $C$ for $t_w=10^5$ and the values of $T$ given in
the legend. The straight lines have slopes $1/T$. FDT holds for
the highest temperature. $T_{\rm eff} \sim 0.08$ in this
$T$-range.}
 \label{fig3}
\end{figure}

It can be easily shown that, in the linear response regime,
\begin{eqnarray}
\delta n(t)= \varphi_0\;\chi(t+t_w,t_w)\;,\;\;
\chi(t,t')=\int_{t'}^{t}\;dt'' R(t,t'')\;, \label{chi}
\end{eqnarray}
where $\chi$ is the local susceptibility and $R$ is the local
response function. At equilibrium, $\chi$ and $C$ are related by
the fluctuation dissipation theorem (FDT), $T\chi(t)=1 - C(t)$. We
show in  Fig.~\ref{fig3}~(a)  a parametric plot of the product
$T\chi$ as a function of $C$ for $T=0.02$ and several waiting
times. It can be seen that our data violate this relationship:
whereas for each $t_w$ the points do align on the FDT straight
line for short times ($C \sim 1$), they deviate from it at long
times (small $C$). The value of $C$ at which deviations first
appear slowly increases with the age of the system.

A generalization of FDT was found to hold for many model
systems~\cite{CuKu} in which, for long $t$ and $t'$, the
off-equilibrium $\chi(t,t')$ still depends on the times only
through its dependence on $C$, $\chi(t,t') =
\Theta\left[C(t,t')\right]$ where $\Theta$ is in general non
linear. In glassy systems below the glass transition the FDT
curves for different values of $t_w$ converge to the limiting
$\Theta\left[C\right]$ as $t_w \to \infty$. Our data are
compatible with $\Theta\left[C\right]$ consisting of two straight
lines with different slopes with a break point at some value
$C^{\star}$ of the correlation. This type of behavior
behavior is found in some mean-field spin-glass
models~\cite{CuKu} and in models of structural
glasses~\cite{parisi} but not in short-range spin glasses.
Since our two-dimensional model has no finite-temperature glass
transition,
 we should see the break point continuously
shift to lower values of $C$ with increasing $t_w$ such
 that for $t_w > \tau_{eq}$ the classical
FDT line is recovered . This process is slow,
however, and our data only show a hint of it: on the time scale of
the simulation the system behaves as a glass that violates FDT. We
display in Fig.~\ref{fig3}~(b) the temperature-dependence of the
FDT diagram. It is seen that the break point gradually shifts to
longer time scales with increasing $T$ until, by the time we reach
$T=0.08$, FDT is fully restored.

In the study of models of glasses it has proven
 conceptually useful to introduce an effective temperature~\cite{Teff}
$1/T_{\rm eff} = -\partial \Theta/\partial C$ that depends on
timescale and represents (loosely speaking) the temperature of the
un-equilibrated degrees of freedom. The data displayed in
Fig.~\ref{fig3}~(b) suggest that at low $T$, in the time-domain
explored, there exist just two timescales, with effective
temperatures $T_{\rm eff} = T$ and $T_{\rm eff} \sim 0.08$,
respectively.

\begin{figure}
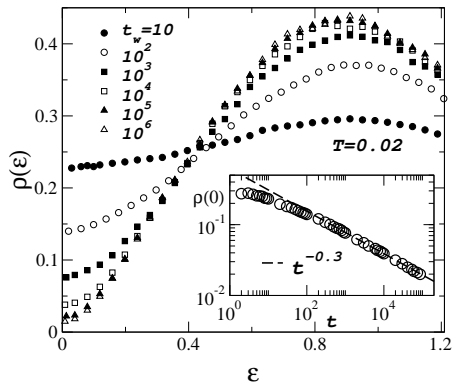

\onefigure[height=5cm]{fig4.eps} \caption{ Time evolution of the
Coulomb gap for $T=0.02$ for the waiting times shown in the
legend. The size of the system is $N=2500$. Inset: power-law decay
of the density of states at the Fermi level. } \label{fig4}
\end{figure}

Slow relaxation can also be seen in
other physical quantities such as  the density of states
 $\rho(\epsilon)=1/N \langle \sum_i \delta(\epsilon-\epsilon_i)\rangle$.
It was suggested~\cite{yu2} that the slow relaxation of the
conductance observed experimentally may reflect a slow formation
of the Efros-Shklovskii Coulomb gap~\cite{coulomb-glass}. We
display in Fig.~(\ref{fig4}) our results for $\rho(\epsilon)$ for
a system of size $N=2500$ at $T=0.02$. It is seen that the density
of states, featureless right after the quench, slowly develops a
pseudo gap for $\left|\epsilon\right| \to 0$. The inset to the
figure shows the time dependence of the density of states at the
Fermi level, $\rho(0,t) \sim t^{-\nu}$ with $\nu \approx 0.3$. A
power-law decay of $\rho(0,t)$ was predicted by the
phenomenological approach of Reference~\cite{yu2}.

The glassy behavior described above only exists on timescales
shorter than $\tau_{eq}$. At fixed $T$, this increases
exponentially with the Coulomb energy which is determined by the
doping, $E_C \propto n^{1/3}$. For a carrier density $n =
10^{21}\;{\rm cm}^{-3}$, $E_C \sim 700 K$ which leads to values of
$\tau_{eq}$ that are much greater than experimental timescales at
the working temperature. The relaxation time decreases very
rapidly with decreasing $n$ implying an absence of glassiness in
the more lightly doped samples, a trend that was observed in the
experiments~\cite{films:aging}. It is interesting to notice that 
2D spin-glass films, which 
like the 2D Coulomb glasses discussed here do not have a finite
temperature phase transition, 
also exhibit aging effects~\cite{eric}. 

In summary, we studied the dynamics of the 2D Coulomb-glass model
at low temperature. The equilibration time of the system diverges
exponentially for $T\to 0$ signaling a transition at $T_{\rm
g}=0$. At finite $T$, the evolution of the system after a sudden
quench from high temperatures is similar to that of glasses below $T_g$
 for times shorter than
$\tau_{\rm eq}$. In this regime the charge autocorrelation
function and the response of to local random
perturbations show simple aging and FDT violations are observed. 
The density of states close to the Fermi level 
evolves slowly and $\rho(0)$
has a power-law decay.

Our results (except those for the response function)
 were obtained in the absence of the symmetry
breaking field $\varphi_i$. We checked explicitly that  
 the aging behavior of $C(t,t')$ persists  
in the presence of a small random local potential
 ($\varphi_0\ll 1$). 
We have preliminary evidence that the model with 
 $\varphi_0 \sim 1$  
 also ages, in contradiction with 
recent claims in the literature~\cite{efros}.    

I thank Z. Ovadyahu, A. Vaknin, M. Pollak,  L. F. Cugliandolo,  A.
Kolton and D. Dominguez for illuminating discussions. This work
was supported in part by NSF Grant No. PHY99-07949 and by the
program ECOS-Sud, project A01E01.


\begin{thebibliography}{99}

\bibitem{davies} Davies J. H. {\it et al.},
Phys. Rev. Lett., {\bf 49} (1982) 758; {\it ibid.}, Phys. Rev. B,
{\bf 29} (1984)  4260.

\bibitem{pollak1} Pollak M., Philos. Mag. B, {\bf 50} (1984) 265;
Gr\"{u}newald  M. {\it et al.}, J. Phys. C, {\bf 15} (1982) L1153.

\bibitem{films:experiments} Ben-Chorin M., {\it et al.},
Phys. Rev. B, {\bf 48} (1993) 15025 ; Ovadyahu  Z.  and Pollak M.,
Phys. Rev. Lett.,{\bf 79} (1997) 459.

\bibitem{films:aging}
Vaknin A. {\it et al.}, Phys. Rev. Lett., {\bf 84} (2000) 3402;
{\it ibid.}, Phys. Rev. B, {\bf 65}  (2002) 134208; {\it ibid},
Phys. Rev. Lett., {\bf 81} (1998) 669.

\bibitem{review:glasses1}
Struik L. C. E., {\it Physical Aging in Amorphous Polymers and
Other Materials} (Elsevier, Amsterdam,1978); Vincent E. {\it et
al}, in {\it Complex behavior of Glassy Systems}, M. Rubi and C.
Perez-Vicente Eds. (Springer, Berlin, 1997).

\bibitem{lee}  Xue W. and  Lee P. A., Phys. Rev. B, {\bf 38}
 (1988) 9093.

\bibitem{yu1} Grannan E. R.  and  Yu Clare C.,
Phys. Rev. Lett., {\bf 71}  (1993) 3335.

\bibitem{rusos} Menashe D.  {\it et
al.}, Europhys. Lett., {\bf 52}  (2000) 94; {\it ibid.}, Phys.
Rev. B, {\bf 64}  (2001) 115209.

\bibitem{efros} Tsigankov  and Efros A. L.,
Phys. Rev. Lett., {\bf 88} (2002) 176602 ;  Tsigankof D. N. {\it
et al.}, Phys. Rev. B, {\bf 68}  (2003) 184205.

\bibitem{coulomb-glass} {\it Electron-Electron interactions
 in Disorder Systems}, A. L. Efros and M. Pollak, Editors
  (North-Holland, Amsterdam, 1985).

\bibitem{bhatt}Bhatt R. N. and Young A. P., Phys. Rev. Lett., {\bf
54} (1984) 924.

\bibitem{rieger} Kisker J. {\it et al.},
Phys. Rev.B, {\bf 53}  (1996) 6418; Berthier L. and Bouchaud
J.-P., Phys. Rev. B, {\bf 66}  (2002) 054404.

\bibitem{godreche}
Godr\`eche C. and Luck J. M., J. Phys. A, {\bf 33}  (2000) 9141.

\bibitem{CuKu} Cugliandolo L. F.  and Kurchan J., Phys. Rev.
Lett., {\bf 71}  (1993) 173; {\it ibid}, J. Phys. A, {\bf 27}
 (1994) 5749.

\bibitem{zvi-new} Vaknin A. {\it et al.}, Phys. Rev. B, {\bf 61}
(2000) 6692; {\it ibid}, Springer Proceedings in Physics, {\bf 87}
 (2001) 995.

\bibitem{Teff} Cugliandolo L. F., Kurchan J. and Peliti L.,
Phys. Rev. E, {\bf 55}  (1997) 3898.

\bibitem{parisi} Parisi G., Phys. Rev. Lett., {\bf 79}  (1997) 3660.

\bibitem{yu2} Yu Clare C.,
Phys. Rev. Lett., {\bf 82} (1999) 4074.

\bibitem{eric} Schins A. G. {\it et al.}, Phys. Rev. B, {\bf 48}  (1993) 16 524.
\end{thebibliography}
\end{document}